\documentclass[aps,prd,superscriptaddress,twoside,twocolumn,nofootinbib,showpacs,floatfix]{revtex4-1}
\usepackage[colorlinks,citecolor=blue]{hyperref}
\usepackage{amsmath,amssymb}
\usepackage{graphicx,bm,braket}
\usepackage{subfigure}
\usepackage{color}
\usepackage{newtxtext, newtxmath}
\pdfpagewidth=\paperwidth
\pdfpageheight=\paperheight
\allowdisplaybreaks
\usepackage{epstopdf}
\newcommand*{\slashed}[1]{{#1\!\!\!/}}
\newcommand*{\hc}{\text{H.\,c.}}

\begin{document}

\title{\boldmath Comprehensive analysis of the $\gamma p \to K^+ \Sigma^0(1385)$, $\gamma n \to K^+ \Sigma^-(1385)$, and $\pi^+ p \to K^+ \Sigma^+(1385)$ reactions}

\author{Ai-Chao Wang}
\affiliation{College of Science, China University of Petroleum (East China), Qingdao 266580, China}

\author{Neng-Chang Wei}
\affiliation{School of Physics, Henan Normal University, Henan 453007, China}

\author{Fei Huang}
\email[Corresponding author. Email: ]{huangfei@ucas.ac.cn}
\affiliation{School of Nuclear Science and Technology, University of Chinese Academy of Sciences, Beijing 101408, China}

\date{\today}

\begin{abstract}
A comprehensive analysis of all available data on cross sections and spin-dependent observables for the $\gamma p \to K^+ \Sigma^0(1385)$, $\gamma n \to K^+ \Sigma^-(1385)$, and $\pi^+ p \to K^+ \Sigma^+(1385)$ reactions is performed within an effective Lagrangian framework. In addition to the $s$-channel $N$ exchange, $t$-channel $K$ and $K^\ast$ exchanges, $u$-channel $\Lambda$, $\Sigma$, and $\Sigma^\ast$ exchanges, and the interaction current, the $s$-channel contributions from a minimal set of $N$ and $\Delta$ resonances are included in constructing the reaction amplitudes to describe the data. The analysis reveals that all available data for both photon- and hadron-induced $K\Sigma(1385)$ production reactions can be simultaneously and satisfactorily reproduced when the contribution of the $s$-channel $\Delta(1930){5/2}^-$ resonance is taken into account. The reaction mechanisms, particularly the resonance contributions in each considered reaction, are thoroughly analyzed and discussed in detail.
\end{abstract}

\pacs{25.20.Lj, 13.60.Le, 14.20.Gk}

\keywords{photoproduction process, effective Lagrangian approach, resonance}

\maketitle

\section{Introduction}   \label{Sec:intro}

The investigation of nucleon resonances ($N^\ast$'s) and $\Delta$ resonances ($\Delta^\ast$'s) remains critically important in hadron physics, providing essential insights into the nonperturbative regime of quantum chromodynamics. Our current understanding of $N^\ast$'s and $\Delta^\ast$'s primarily stems from $\pi N$ scattering or $\pi$ and $K$ photoproduction reactions. However, quark models \cite{Isgur:1978,Capstick:1986,Loring:2001} predict a significant greater number of $N^\ast$ and $\Delta^\ast$ states than experimentally observed. A plausible explanation for this discrepancy is that some $N^\ast$'s and $\Delta^\ast$'s couple weakly to $\pi N$, $K\Lambda$, or $K\Sigma$ final states but strongly to other meson production reactions. Therefore, it becomes both interesting and imperative to explore $N^\ast$'s and $\Delta^\ast$'s in production reactions other than $\pi N$, $K\Lambda$, and $K\Sigma$. 

In this study, we focus on a combined analysis of the $K\Sigma(1385)$ production reactions for which the experimental data are available: $\gamma p \to K^+ \Sigma^0(1385)$, $\gamma n \to K^+ \Sigma^-(1385)$, and $\pi^+ p \to K^+ \Sigma^+(1385)$. The higher threshold of $K\Sigma(1385)$ makes these reactions particularly well-suited for investigating $N^\ast$'s and $\Delta^\ast$'s in the relatively less-explored high-energy region. Furthermore, and most importantly, a combined analysis of all available data for both photon- and hadron-induced $K\Sigma(1385)$ production reactions can impose stronger constrains on the theoretical model, leading to significantly more reliable and robust results.

Experimentally, high-precision data on differential cross sections and total cross sections for the $\gamma p\to K^+\Sigma^0(1385)$ reaction in the center-of-mass energy range $W \approx 2.0$–$2.8$ GeV were reported by the CLAS Collaboration in 2013 \cite{Mori:2013}. For the $\gamma n\to K^+\Sigma^-(1385)$ reaction, high-statistics data on differential cross sections and photobeam asymmetries in the photon laboratory incident energy range $E_\gamma = 1.5$–$2.4$ GeV were published by the LEPS Collaboration in 2009 \cite{Hicks:2009}. Subsequently, in 2014, the CLAS Collaboration reported preliminary differential cross-section data for the same reaction in the photon laboratory incident energy range $E_\gamma = 1.6$–$2.4$ GeV \cite{Paul:2014}, covering a broader range of scattering angles compared to the LEPS data. As for the $\pi^+ p \to K^+ \Sigma^+(1385)$ reaction, the only available data are the differential cross sections measured in the center-of-mass energy range $W \approx 1.9$–$2.1$ GeV, published in 1971 \cite{Hanson:1972}.

In our previous work, Ref.~\cite{Wang:2020}, we investigated the $\gamma p \to K^+\Sigma^0(1385)$ reaction within an effective Lagrangian approach. There, the reaction amplitudes were constructed by including the $s$-channel nucleon ($N$) exchange, the $t$-channel $K$ and $K^\ast$ exchanges, the $u$-channel $\Lambda$ exchange, the interaction current, and the $s$-channel contributions from a minimum set of $N$ and $\Delta$ resonances. It was demonstrated that the high-precision differential cross-section data from the CLAS Collaboration for $\gamma p \to K^+\Sigma^0(1385)$ could be well reproduced by including one of the following resonances in the $s$ channel: $N(1895){1/2}^-$, $\Delta(1900){1/2}^-$, or $\Delta(1930){5/2}^-$. Subsequently, in Ref.~\cite{Wang:2022}, we extended our study to the $\gamma n\to K^+\Sigma^-(1385)$ reaction using the same theoretical framework developed in Ref.~\cite{Wang:2020} for the $\gamma p \to K^+\Sigma^0(1385)$ reaction. However, due to the electric charge constraints of the initial and final hadrons, the $u$-channel $\Lambda$ exchange was replaced with $\Sigma^\ast$ exchange. It was found that the data on differential cross sections and photo-beam asymmetries for $\gamma n\to K^+\Sigma^-(1385)$ could be well described by including the $\Delta(1930){5/2}^-$ resonance, but not the $N(1895){1/2}^-$ or $\Delta(1900){1/2}^-$ resonances. Note that the cutoff parameter for the $t$-channel $K$ exchange and the coupling constants for the $s$-channel $\Delta(1930){5/2}^-$ resonance should, in principle, be consistent across both the $\gamma p \to K^+\Sigma^0(1385)$ and $\gamma n\to K^+\Sigma^-(1385)$ reactions. However, in Ref.~\cite{Wang:2022}, these parameters were refitted rather than adopted from Ref.~\cite{Wang:2020} to better reproduce the $\gamma n\to K^+\Sigma^-(1385)$ data.

In the present work, we conduct a combined analysis of all available data on cross sections and spin-dependent observables for the $\gamma p \to K^+ \Sigma^0(1385)$, $\gamma n \to K^+ \Sigma^-(1385)$, and $\pi^+ p \to K^+ \Sigma^+(1385)$ reactions within an effective Lagrangian approach. This study differs from our previous works in Refs.~\cite{Wang:2020,Wang:2022} mainly in three key aspects. First, the theoretical framework employed here is self-consistent for all those three reactions. Specifically, the hadronic coupling constants, cutoff parameters, and electromagnetic couplings for isospin-$3/2$ resonances are uniquely determined for the same vertices appearing in these three reactions. Second, the differential cross-section data for the $\pi^+ p \to K^+ \Sigma^+(1385)$ reaction are now included in the database, providing additional constrains on the theoretical model. Finally, a thorough and systematic analysis of the contributions of $N$ and $\Delta$ resonances in these reactions is performed by testing all possible resonance candidates within the considered energy region. Owing to these significant  improvements, the present work is expected to yield a deeper understanding of the reaction mechanisms for $\gamma p \to K^+ \Sigma^0(1385)$, $\gamma n \to K^+ \Sigma^-(1385)$, and $\pi^+ p \to K^+ \Sigma^+(1385)$, as well as a more reliable extraction of resonance properties and parameters consistent with the available experimental data.  

The paper is organized as follows. In Sec.~\ref{Sec:formalism}, we present a concise outline of the theoretical framework used in our model. In Sec.~\ref{Sec:results}, we discuss the theoretical results in detail, accompanied by an in-depth analysis. Finally, Sec.~\ref{Sec:summary} provides a summary of our findings.

\section{Formalism}  \label{Sec:formalism}

\begin{figure}[tbp]
\subfigure[~$s$ channel]{
\includegraphics[width=0.45\columnwidth]{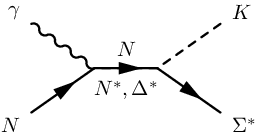}}  {\hglue 0.4cm}
\subfigure[~$t$ channel]{
\includegraphics[width=0.45\columnwidth]{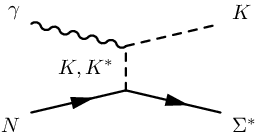}} \\[6pt]
\subfigure[~$u$ channel]{
\includegraphics[width=0.45\columnwidth]{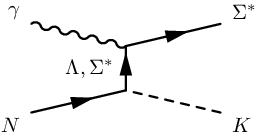}} {\hglue 0.4cm}
\subfigure[~Interaction current]{
\includegraphics[width=0.45\columnwidth]{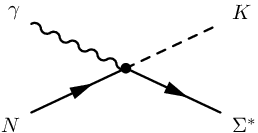}}
\caption{Generic structure of the amplitudes for $\gamma p\to K^+ \Sigma^0(1385)$ and $\gamma n\to K^+ \Sigma^-(1385)$. Time proceeds from left to right. The symbols $\Sigma^\ast$ and $K^\ast$ denote $\Sigma(1385)$ and $K^\ast(892)$, respectively.  }
\label{FIG:feymans-g}
\end{figure}

\begin{figure}[tbp]
\subfigure[~$s$ channel]{
\includegraphics[width=0.46\columnwidth]{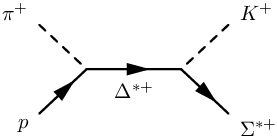}}  {\hglue 0.3cm}
\subfigure[~$t$ channel]{
\includegraphics[width=0.46\columnwidth]{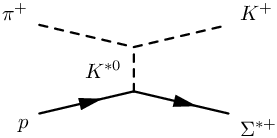}} \\[14pt]
\subfigure[~$u$ channel]{
\includegraphics[width=0.46\columnwidth]{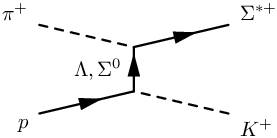}}
\caption{Same as Fig.~\ref{FIG:feymans-g} but for $\pi^+ p \to K^+ \Sigma^+(1385)$.  }
\label{FIG:feymans}
\end{figure}

In this study, we present a comprehensive analysis of the $\gamma p \to K^+ \Sigma^0(1385)$, $\gamma n \to K^+ \Sigma^-(1385)$, and $\pi^+ p \to K^+ \Sigma^+(1385)$ reactions within an effective Lagrangian model. For the $\gamma p \to K^+ \Sigma^0(1385)$ and $\gamma n \to K^+ \Sigma^-(1385)$ reactions, following our previous works in Refs.~\cite{Wang:2020,Wang:2022}, the following contributions as depicted in Fig.~\ref{FIG:feymans-g} are considered in constructing the reaction amplitudes: (i) $s$-channel $N$, $N^\ast$, and $\Delta^\ast$ exchanges, (ii) $t$-channel $K$ and $K^\ast(892)$ exchanges, (iii) $u$ channel $\Lambda$ and $\Sigma^\ast$ exchanges, and (iv) an interaction current. The interaction current is introduced in such a way that the full photoproduction amplitudes satisfy the generalized Ward-Takahashi Identity (gWTI), thereby guaranteeing gauge invariance \cite{Haberzettl:1997}. The $\Sigma$ exchange in the $u$-channel is not considered, since we don't have experimental information on the $\Sigma^\ast \Sigma \gamma$ coupling, and moreover, this diagram is expected to contribute considerably only at high energy backward angles where the data for $\gamma p \to K^+ \Sigma^0(1385)$ and $\gamma n \to K^+ \Sigma^-(1385)$ are scarce. Note that in $\gamma n \to K^+ \Sigma^-(1385)$, the $u$-channel $\Lambda$ exchange does not contribute due to electric charge conservation. For the $\pi^+ p \to K^+ \Sigma^+(1385)$ reaction, we consider the following contributions, as illustrated in Fig.~\ref{FIG:feymans}, to construct the $s$-, $t$-, and $u$-channel amplitudes: (i) $s$-channel $\Delta^{\ast +}$ exchange, (ii) $t$-channel $K^{\ast 0}(892)$ exchange, and (iii) $u$-channel $\Lambda$ and $\Sigma^0$ exchanges. 

By use of the effective Lagrangians given in Sec.~\ref{Sec:Lagrangians}, the $s$-, $t$-, and $u$-channel amplitudes can be constructed straightforwardly by evaluating the corresponding Feynman diagrams as illustrated in Figs.~\ref{FIG:feymans-g}-\ref{FIG:feymans}. The interaction current for $\gamma p \to K^+ \Sigma^0(1385)$ and $\gamma n \to K^+ \Sigma^-(1385)$ is modeled by a generalized contact current as done in Refs.~\cite{Haberzettl:2006,Huang:2012,Huang:2013}:
\begin{equation}
{\cal M}^{\nu\mu}_{\rm int} = \Gamma^\nu_{\Sigma^\ast N K}(q) C^\mu + {\cal M}^{\nu\mu}_{\rm KR} f_t.  \label{eq:Mint}
\end{equation}
Here $\Gamma^\nu_{\Sigma^\ast NK}(q)$ is the $\Sigma(1385) NK$ vertex function,
\begin{equation}
\Gamma^\nu_{\Sigma^\ast NK}(q) = - \frac{g_{\Sigma^\ast NK}}{M_K} q^\nu,
\end{equation}
with $q$ and $M_K$ being the four-momentum and mass of the outgoing $K$ meson, respectively. This vertex function is depicted by the Lagrangian of $\Sigma(1385) NK$ coupling,
\begin{equation}
{\cal L}_{\Sigma^\ast NK} = \frac{g_{\Sigma^\ast NK}}{M_K} \bar{\Sigma}^{\ast \mu} \left( \partial_\mu K \right) N  + \hc,    \label{eq:L_KNSt}
\end{equation}
where the coupling constant $g_{\Sigma^\ast NK}$ is fixed by the SU(3) flavor symmetry relation,
\begin{equation}
\frac{g_{\Sigma^\ast NK}}{M_K} = - \frac{1}{\sqrt{6}}\frac{g_{\Delta N\pi }}{M_\pi},  \label{g_SigstNK}
\end{equation}
with the empirical value $g_{\Delta N\pi}=2.23$ being employed, which results in $g_{\Sigma^\ast NK}=-3.22$. $C^\mu$ in Eq.~\eqref{eq:Mint} is an auxiliary current which is nonsingular and introduced to ensure that the full photoproduction amplitudes are fully gauge invariant. One of the possible choices of $C^\mu$ is \cite{Haberzettl:2006,Huang:2012,Huang:2013}
\begin{align}
C^\mu = & - Q_K \tau_t \frac{f_t - \hat{F}}{t - q^2} \left(2q-k\right)^\mu - Q_{\Sigma^\ast} \tau_u \frac{f_u - \hat{F}}{u - p'^2} \left(2p'-k\right)^\mu  \nonumber \\
             & - \tau_s Q_N \frac{f_s - \hat{F}}{s - p^2} \left(2p+k\right)^\mu,   \label{eq:Cmu}
\end{align}
with
\begin{align}
\hat{F} = 1 - \hat{h} \left(1-\delta_s f_s\right) \left(1-\delta_u f_u\right) \left(1-\delta_t f_t\right),   \label{eq:Fhat}
\end{align}
where $s$, $t$, and $u$ are the Mandelstam variables of the internally exchanged particles; the constant $\delta_x$ $(x=s,u,t)$ is unity if the corresponding $x$-channel contributes and zero otherwise; $Q_K$, $Q_{\Sigma^\ast}$, and $Q_N$ are electric charges of the outgoing $K$, $\Sigma(1385)$, and incoming $N$, respectively; $\tau_x$ $(x=s,u,t)$ denotes the isospin factor of the corresponding $x$-channel hadronic vertex; $p$, $p'$, and $k$ are four-momenta of the incoming $N$, outgoing $\Sigma(1385)$, incoming photon, respectively; and $\hat{h}$ is an arbitrary function that goes to unity in high-energy limit to prevent the violation of scaling behavior as denoted in Ref.~\cite{Drell:1972}. In the present work, $\hat{h}$ is parametrized as
\begin{equation}
\hat{h} = 1 - A_{0}\frac{\Lambda_{c}^{4}}{\Lambda_{c}^{4}+\left(s-s_{\rm th}\right)^2},  \label{eq:A0}
\end{equation}
with
\begin{equation}
s_{\rm th}=\left(M_{\Sigma^\ast} + M_K\right)^2,
\end{equation}
where $M_{\Sigma^\ast}$ is the mass of $\Sigma(1385)$ baryon; the cutoff $\Lambda_c$ is fixed to be $2.5$ GeV to make $\hat{h}$ not rise too rapidly for the energy range considered; and the strength $A_0$ is treated as a fit parameter. The $f_s$, $f_u$, and $f_t$ in Eqs.~\eqref{eq:Mint}, \eqref{eq:Cmu}, and \eqref{eq:Fhat} are form factors for $s$-channel $N$ exchange, $u$-channel $\Sigma^*$ exchange, and $t$-channel $K$ exchange, respectively, with their explicit forms being given in Sec.~\ref{Sec:form_factor}. The Kroll-Ruderman term ${\cal M}_{\rm KR}^{\nu\mu}$ in Eq.~\eqref{eq:Mint} reads
\begin{equation}
{\cal M}^{\nu\mu}_{\rm KR} = \frac{g_{\Sigma^\ast NK}}{M_K}  g^{\nu\mu} \tau Q_K,
\end{equation}
which is resulted from the effective Lagrangian of $\Sigma(1385) NK\gamma$ coupling,
\begin{equation}
{\cal L}_{\Sigma^\ast N K\gamma } = -iQ_K \frac{g_{\Sigma^\ast NK}}{M_K} \bar{\Sigma}^{\ast \mu} A_\mu K N  + \hc,   \label{eq:L_rKNSt}
\end{equation}
obtained by the minimal gauge substitution $\partial_\mu \to {\cal D}_\mu\equiv \partial_\mu - i Q_K A_\mu$ in the $\Sigma^\ast NK$ interaction Lagrangian of Eq.~(\ref{eq:L_KNSt}). 

In the remainder of this section, we present the effective Lagrangians, propagators, and form factors used to calculate the $s$-, $t$-, and $u$-channel amplitudes for the $\gamma p \to K^+ \Sigma^0(1385)$, $\gamma n \to K^+ \Sigma^-(1385)$, and $\pi^+ p \to K^+ \Sigma^+(1385)$ reactions. The formulae for the $\gamma p \to K^+ \Sigma^0(1385)$ and $\gamma n \to K^+ \Sigma^-(1385)$ reactions are consistent with those outlined in our previous works \cite{Wang:2020,Wang:2022}.

\subsection{Effective Lagrangians} \label{Sec:Lagrangians}

The effective interaction Lagrangians used in the present work are given below. For further convenience, we define the field
\begin{equation}
\Sigma^\ast = \Sigma(1385),
\end{equation}
the operators
\begin{equation}
\Gamma^{(+)}=\gamma_5  \quad {\rm{and}} \quad  \Gamma^{(-)}=1,
\end{equation}
and the field-strength tensors
\begin{equation}
F^{\mu\nu} = \partial^{\mu}A^\nu-\partial^{\nu}A^\mu,
\end{equation}
with $A^\mu$ denoting the electromagnetic field.

The hadronic interaction Lagrangians, in addition to the one given in Eq.~\eqref{eq:L_KNSt}, required to calculate the non-resonant Feynman diagrams as shown in Figs.~\ref{FIG:feymans-g}-\ref{FIG:feymans} are 
\begin{align}
{\cal L}_{\Sigma^\ast N K^\ast} &= - i \frac{g_{\Sigma^\ast N K^\ast}}{2M_N}{\bar\Sigma}^\ast_\mu \gamma_\nu \gamma_5 {K^\ast}^{\mu \nu}N  + \hc,  \\[6pt]
{\cal L}_{\Lambda NK} &= - \frac{g_{\Lambda NK}}{2M_N}\bar{\Lambda} \gamma_5 \gamma^\mu \left( \partial_\mu K \right) N + \hc,  \label{eq:L_LNK}    \\[6pt]
{\cal L}_{\Sigma NK} &= - \frac{g_{\Sigma NK}}{2M_N}\bar{\Sigma} \gamma_5 \gamma^\mu \left( \partial_\mu K \right) N + \hc,  \label{eq:L_SNK}    \\[6pt]
{\cal L}_{K^\ast K \pi} &=  i g_{K^\ast K \pi} \left[ \left( \partial_\mu K^\dagger K^{\ast\mu} \right)\cdot \pi - \left( K^\dagger K^{\ast\mu} \right) \cdot \partial_\mu \pi \right],  \label{Lag:kstarkpi}  \\[6pt]
{\cal L}_{\Sigma^\ast \Lambda \pi} &= \frac{g_{\Sigma^\ast \Lambda \pi}}{M_\pi} \left(\bar{\Sigma}^{\ast\mu}\cdot \partial_\mu\pi\right) \Lambda + \hc, \label{eq:L_SLP}    \\[6pt]
{\cal L}_{\Sigma^\ast \Sigma \pi} &= - i \frac{g_{\Sigma^\ast \Sigma \pi}}{M_\pi} \left(\bar{\Sigma}^{\ast\mu}\times \Sigma \right) \cdot \partial_\mu \pi + \hc,  \label{eq:L_SSP}
\end{align}
with $M_N$ and $M_\pi$ being the masses of $N$ baryon and $\pi$ meson, respectively. The coupling constant $g_{\Sigma^\ast NK^\ast}$, $g_{\Lambda NK}$, and $g_{\Sigma NK}$ are fixed by the SU(3) flavor symmetry relations,
\begin{align}
g_{\Sigma^\ast NK^\ast} &= -\frac{1}{\sqrt{6}} g_{\Delta N\rho},   \label{g_SigstNKst} \\[6pt]
g_{\Lambda NK} &= -\frac{3\sqrt{3}}{5} g_{NN\pi},   \label{g_LamNK} \\[6pt]
g_{\Sigma NK} &= \frac{1}{5} g_{NN\pi},
\end{align}
with the empirical values $g_{\Delta N\rho}=-39.1$ and $g_{NN\pi} = 13.46$ being utilized \cite{Huang:2012}. The coupling constants $g_{K^\ast K \pi}=3.24$, $g_{\Sigma^\ast \Lambda \pi}=1.26$, $g_{\Sigma^\ast \Sigma \pi}=0.98$ are determined from the partial decay widths ${\Gamma}_{K^\ast \to K \pi} \simeq 47.3$ MeV, ${\Gamma}_{\Sigma^\ast \to \Lambda \pi} \simeq 36 \times 87.0\% = 31.32$ MeV, and ${\Gamma}_{\Sigma^\ast \to \Lambda \pi} \simeq 36 \times 11.7\% = 4.21$ MeV, respectively, which are quoted from the most recent Review of Particle Physics (RPP) \cite{PDG:2024}.

The electromagnetic interaction Lagrangians, in addition to the one given in Eq.~\eqref{eq:L_rKNSt}, required to calculate the non-resonant Feynman diagrams as shown in Figs.~\ref{FIG:feymans-g}-\ref{FIG:feymans} are 
\begin{align}
{\cal L}_{NN\gamma} =& - e \bar{N} \left[ \left( \hat{e} \gamma^\mu - \frac{\hat{\kappa}_N}{2M_N}\sigma^{\mu \nu}\partial_{\nu} \right) A_\mu \right] N,  \\[6pt]
{\cal L}_{\gamma K{K^\ast}} =& \, e\frac{g_{\gamma K{K^\ast}}}{M_K}\varepsilon^{\alpha \mu \lambda \nu}\left(\partial_\alpha A_\mu\right)\left(\partial_\lambda K\right)K^\ast_\nu, \label{Lag:gKKst} \\[6pt]
{\cal L}_{\gamma KK} =& \, i e \left[ K^+ \left(  \partial^\mu K^- \right)  -  K^- \left(\partial^\mu K^+ \right) \right] A_\mu,  \\[6pt]
{\cal L}_{\Sigma^\ast \Lambda \gamma} =& -ie\frac{g^{(1)}_{\Sigma^\ast \Lambda \gamma}}{2M_N} \bar{\Sigma}^\ast_\mu \gamma_\nu \gamma_5 F^{\mu \nu} \Lambda \nonumber \\
& +e\frac{g^{(2)}_{\Sigma^\ast \Lambda \gamma}}{\left(2M_N\right)^2} \bar{\Sigma}^\ast_\mu \gamma_5 F^{\mu \nu}\partial_\nu \Lambda  + \hc,   \\[6pt]
\mathcal{L}_{\gamma \Sigma^\ast\Sigma^\ast} =& \, Q_{\Sigma^\ast} \bar{\Sigma}^\ast_\mu A_{\alpha} \left[g^{\mu\nu}\gamma^{\alpha} - \frac{1}{2} \left\{\gamma^{\mu}\gamma^{\nu}, \gamma^{\alpha} \right\} \right] \Sigma^\ast_{\nu}  \nonumber \\
& -e\frac{\kappa_{\Sigma^\ast}}{2M_N} \bar{\Sigma}^\ast_\mu \sigma^{\alpha\beta}\left(\partial_{\beta} A_{\alpha}\right) g^{\mu\nu} \Sigma^\ast_{\nu} ,
\end{align}
where $e$ is the elementary charge unit; $\hat{e}$ stands for the charge operator; $\hat{\kappa}_N = \kappa_p \left(1+\tau_3\right)/2 + \kappa_n \left(1-\tau_3\right)/2$ with the anomalous magnetic moments $\kappa_p=1.793$ and $\kappa_n=-1.913$; $\varepsilon^{\alpha \mu \lambda \nu}$ is the totally antisymmetric Levi-Civita tensor with $\varepsilon^{0123}=1$; and $\kappa_{\Sigma^\ast}$ denotes the anomalous magnetic moment of $\Sigma(1385)$ taken as $\kappa_{\Sigma^{\ast-}}=-2.43$ from a quark model prediction \cite{Lichtenberg:1977}. The coupling constant $g_{\gamma K K^\ast}$ is determined by fitting the radiative decay width of $K^\ast(892) \to K\gamma$ given by the RPP \cite{PDG:2024}, which leads to $g_{\gamma K^\pm K^{\ast\pm}}=0.413$ with the sign inferred from $g_{\gamma \pi \rho}$ \cite{Garcilazo:1993} via the flavor SU(3) symmetry considerations in conjunction with the vector-meson dominance assumption. The coupling constants $g^{(1)}_{\Sigma^\ast \Lambda \gamma}$ and $g^{(2)}_{\Sigma^\ast \Lambda \gamma}$ are constrained by the decay width of $\Gamma_{\Sigma^0(1385)\to\Lambda\gamma}=0.45$ MeV \cite{PDG:2024}, which means only one of them is free. In practice, we treat the ratio of them as a fit parameter.

The hadronic interaction Lagrangians required to calculate the resonant Feynman diagrams as shown in Figs.~\ref{FIG:feymans-g}-\ref{FIG:feymans} are 
\begin{align}
\mathcal{L}_{R\Sigma^\ast K}^{1/2\pm} =& \frac{g_{R\Sigma^\ast K}^{(1)}}{M_K} \bar{\Sigma}^{\ast}_\mu \Gamma^{(\mp)} \left(\partial^\mu K\right) R + \hc,   \\[6pt]
\mathcal{L}^{1/2 \pm}_{RN\pi} =& - \frac{g_{RN\pi}}{2M_N} {\bar N} \Gamma^{(\pm)} \left(\slashed{\partial}\pi \right) R + \hc,   \\[6pt]
\mathcal{L}_{R\Sigma^\ast K}^{3/2\pm} =& \, \frac{g_{R\Sigma^\ast K}^{(1)}}{M_K} \bar{\Sigma}^{\ast}_\mu  \gamma^\alpha \Gamma^{(\pm)} \left(\partial_\alpha K\right) R^\mu \nonumber \\
& + i \frac{g_{R\Sigma^\ast K}^{(2)}}{M_K^2} \bar{\Sigma}^\ast_\alpha \Gamma^{(\pm)} \left(\partial^\mu \partial^\alpha K\right)  R_\mu + \hc,   \\[6pt]
\mathcal{L}^{3/2\pm}_{RN\pi} =& \, \frac{g_{RN\pi}}{M_\pi} {\bar{R}^\mu} \Gamma^{(\mp)} \left( \partial_\mu \pi \right) N + \hc,     \label{3hfNP}    \\[6pt]
\mathcal{L}_{R\Sigma^\ast K}^{5/2\pm} =& \, i \frac{ g_{R\Sigma^\ast K}^{(1)} }{M_K^2} \bar{\Sigma}^{\ast}_\alpha \gamma^\mu \Gamma^{(\mp)} \left(\partial_\mu \partial_\beta K\right)  R^{\alpha\beta}\nonumber \\
&  - \frac{ g_{R\Sigma^\ast K}^{(2)} }{M_K^3} \bar{\Sigma}^\ast_\mu \Gamma^{(\mp)} \left( \partial^\mu \partial^\alpha \partial^\beta K \right)  R_{\alpha\beta} + \hc,  \\[6pt]
\mathcal{L}^{5/2\pm}_{RN\pi} =& \,  i\frac{g_{RN\pi}}{M_\pi^2} \bar{N} \Gamma^{(\pm)} \left(\partial^\mu\partial^\nu \pi \right) R_{\mu\nu} + \hc,      \label{5hfNP}       \\[6pt]
\mathcal{L}_{R\Sigma^\ast K}^{7/2\pm} =& - \frac{ g_{R\Sigma^\ast K}^{(1)} }{M_K^3} \bar{\Sigma}^{\ast}_\alpha \gamma^\mu \Gamma^{(\pm)} \left(\partial_\mu \partial_\beta \partial_\lambda K\right)  R^{\alpha\beta\lambda}  \nonumber \\
& - i \frac{ g_{R\Sigma^\ast K}^{(2)} }{M_K^4} \bar{\Sigma}^\ast_\mu \Gamma^{(\pm)} \left( \partial^\mu \partial^\alpha \partial^\beta \partial^\lambda K \right)  R_{\alpha\beta\lambda}  \nonumber \\
& + \hc,   \\[6pt]
\mathcal{L}^{7/2\pm}_{RN\pi} =& - \frac{g_{RN\pi}}{M_\pi^3} \bar{R}_{\mu\nu\alpha} \Gamma^{(\mp)}
\left(\partial^\mu\partial^\nu\partial^\alpha \pi \right) N + \hc      \label{7hfNP}
\end{align}
where $R$ designates the $N$ or $\Delta$ resonance, and the superscripts of $\mathcal{L}_{R\Sigma^\ast K}$ and $\mathcal{L}_{RN\pi}$ denote the spin and parity of the resonance $R$. In the current study, the resonance and $N\pi$ coupling constant, $g_{RN\pi}$, is constrained by the resonance's partial decay width $\Gamma_{R\to N\pi}$ advocated in RPP \cite{PDG:2024}. The resonance and $\Sigma^\ast K$ coupling constants, $g_{R\Sigma^\ast K}^{(i)}$ $(i=1,2)$, are treated as free parameters to be determined by fitting the data. 

The electromagnetic interaction Lagrangians required to calculate the resonant Feynman diagrams as shown in Fig.~\ref{FIG:feymans-g} are 
\begin{align}
{\cal L}_{RN\gamma}^{1/2\pm} =& \, e\frac{g_{RN\gamma}^{(1)}}{2M_N}\bar{R} \Gamma^{(\mp)}\sigma_{\mu\nu} \left(\partial^\nu A^\mu \right) N  + \hc, \\[6pt]
{\cal L}_{RN\gamma}^{3/2\pm} =& - ie\frac{g_{RN\gamma}^{(1)}}{2M_N}\bar{R}_\mu \gamma_\nu \Gamma^{(\pm)}F^{\mu\nu}N \nonumber \\
&  + e\frac{g_{RN\gamma}^{(2)}}{\left(2M_N\right)^2}\bar{R}_\mu \Gamma^{(\pm)}F^{\mu \nu}\partial_\nu N + \hc, \\[6pt]
{\cal L}_{RN\gamma}^{5/2\pm} =& \, e\frac{g_{RN\gamma}^{(1)}}{\left(2M_N\right)^2}\bar{R}_{\mu \alpha}\gamma_\nu \Gamma^{(\mp)}\left(\partial^{\alpha} F^{\mu \nu}\right)N \nonumber \\
& \pm ie\frac{g_{RN\gamma}^{(2)}}{\left(2M_N\right)^3}\bar{R}_{\mu \alpha} \Gamma^{(\mp)}\left(\partial^\alpha F^{\mu \nu}\right)\partial_\nu N + \hc, \\[6pt]
{\cal L}_{RN\gamma}^{7/2\pm} =& \, ie\frac{g_{RN\gamma}^{(1)}}{\left(2M_N\right)^3}\bar{R}_{\mu \alpha \beta}\gamma_\nu \Gamma^{(\pm)}\left(\partial^{\alpha}\partial^{\beta} F^{\mu \nu}\right)N \nonumber \\
&  - e\frac{g_{RN\gamma}^{(2)}}{\left(2M_N\right)^4}\bar{R}_{\mu \alpha \beta} \Gamma^{(\pm)} \left(\partial^\alpha \partial^\beta F^{\mu \nu}\right) \partial_\nu N  \nonumber \\
& + \hc,
\end{align}
where the coupling constants are treated as free parameters to be fixed by fitting the data.

\subsection{Propagators}

For $t$-channel $K$ and $K^\ast$ exchanges, the propagators are denoted as
\begin{align}
S_K &= \frac{i}{t-M_K^2},  \\[6pt]
S_{K^\ast} &= \frac{i}{t-M_{K^\ast}^2} \left( -g^{\mu\nu} + \frac{q_t^{\mu}q_t^{\nu}}{M_{K^\ast}^2} \right),
\end{align}
with $q_t$ being the four momentum of the intermediate $K^\ast$ meson.
 
For $u$-channel $\Lambda$, $\Sigma$, and $\Sigma^\ast$ exchanges, the propagators are represented as
\begin{equation}
S_{\Lambda(\Sigma,\Sigma^\ast)} = \frac{i}{\slashed{p}_u - M_{\Lambda(\Sigma,\Sigma^\ast)}},
\end{equation}
with $p_u$ being the four momentum of the intermediate $\Lambda$ $(\Sigma,\Sigma^\ast)$ baryon.

For $s$-channel $N$ exchange, the propagator reads
\begin{equation}
S_N = \frac{i}{\slashed{p}_s - M_N},
\end{equation}
with $p_s$ being the four momentum of the intermediate $N$ baryon.

For $s$-channel exchanges of resonances with spin $1/2$, $3/2$, $5/2$, and $7/2$, the propagators are prescribed as \cite{Behrends:1957,Fronsdal:1958,Zhu:1999}
\begin{align}
S_{1/2} =& \, \frac{i}{\slashed{p}_s - M_R + i \Gamma_R/2},  \\[6pt]
S_{3/2} =& \, \frac{i}{\slashed{p}_s - M_R + i \Gamma_R/2} \left( \tilde{g}^{\mu \nu} + \frac{1}{3} \tilde{\gamma}^\mu \tilde{\gamma}^\nu \right),  \\[6pt]
S_{5/2} =& \, \frac{i}{\slashed{p}_s - M_R + i \Gamma_R/2} \,\bigg[ \, \frac{1}{2} \big(\tilde{g}^{\mu \alpha} \tilde{g}^{\nu \beta} + \tilde{g}^{\mu \beta} \tilde{g}^{\nu \alpha} \big)  \nonumber \\
& - \frac{1}{5}\tilde{g}^{\mu \nu}\tilde{g}^{\alpha \beta}  + \frac{1}{10} \big(\tilde{g}^{\mu \alpha}\tilde{\gamma}^{\nu} \tilde{\gamma}^{\beta} + \tilde{g}^{\mu \beta}\tilde{\gamma}^{\nu} \tilde{\gamma}^{\alpha}  \nonumber \\
& + \tilde{g}^{\nu \alpha}\tilde{\gamma}^{\mu} \tilde{\gamma}^{\beta} +\tilde{g}^{\nu \beta}\tilde{\gamma}^{\mu} \tilde{\gamma}^{\alpha} \big) \bigg],   \\[6pt]
S_{7/2} =& \, \frac{i}{\slashed{p}_s - M_R + i \Gamma_R/2} \, \frac{1}{36}\sum_{P_{\mu} P_{\nu}} \bigg( \tilde{g}^{\mu_1 \nu_1}\tilde{g}^{\mu_2 \nu_2}\tilde{g}^{\mu_3 \nu_3} \nonumber \\
& - \frac{3}{7}\tilde{g}^{\mu_1 \mu_2}\tilde{g}^{\nu_1 \nu_2}\tilde{g}^{\mu_3 \nu_3} + \frac{3}{7}\tilde{\gamma}^{\mu_1} \tilde{\gamma}^{\nu_1} \tilde{g}^{\mu_2 \nu_2}\tilde{g}^{\mu_3 \nu_3} \nonumber \\
& - \frac{3}{35}\tilde{\gamma}^{\mu_1} \tilde{\gamma}^{\nu_1} \tilde{g}^{\mu_2 \mu_3}\tilde{g}^{\nu_2 \nu_3} \bigg),  \label{propagator-7hf}
\end{align}
where
\begin{align}
\tilde{g}^{\mu \nu} =& - g^{\mu \nu} + \frac{p_s^{\mu} p_s^{\nu}}{M_R^2}, \\[6pt]
\tilde{\gamma}^{\mu} =& \, \gamma_{\nu} \tilde{g}^{\nu \mu} = -\gamma^{\mu} + \frac{p_s^{\mu}\slashed{p}_s}{M_R^2},
\end{align}
and $M_R$, $\Gamma_R$, and $p_s$ are, respectively, the mass, width, and four momentum of the resonance $R$. The summation over $P_{\mu}(P_{\nu})$ in Eq.~(\ref{propagator-7hf}) goes over the $3! = 6$ possible permutations of the indices $\mu_1\mu_2\mu_3(\nu_1\nu_2\nu_3)$.

\subsection{Form factors}  \label{Sec:form_factor}

In effective Lagrangian approaches, phenomenological form factors are typically introduced at hadronic vertices to regularize the reaction amplitudes and partially account for the internal structures of hadrons. Following Refs.~\cite{Wang:2017,Wang:2018,Kim:2014,Kim:2011}, we adopt the form factor for intermediate baryon exchange as
\begin{equation}
f_B(p^2) = \left(\frac{\Lambda_B^4}{\Lambda_B^4+\left(p^2-M_B^2\right)^2}\right)^2,  \label{eq:ff_B}
\end{equation}
where $p$, $M_B$, and $\Lambda_B$ denote the four momentum, mass, and cutoff mass of the exchanged baryon $B$, respectively, with the last being considered as a fit parameter. For intermediate meson exchange, we employ the form factor
\begin{equation}
f_M(q^2) = \left(\frac{\Lambda_M^2-M_M^2}{\Lambda_M^2-q^2}\right)^2,     \label{eq:ff_M}
\end{equation}
where $q$, $M_M$, and $\Lambda_M$ represent the four momentum, mass, and cutoff mass of the intermediate meson, respectively, with the last being treated as a fit parameter. 

In this study, we utilize the same cutoff parameter $\Lambda_u$ for all $u$-channel baryon exchanges. Likewise, a single cutoff parameter $\Lambda_s$ is applied to all baryon exchanges in $s$ channel, and $\Lambda_t$ is employed uniformly for all meson exchanges in $t$ channel.

\section{Results and discussion}   \label{Sec:results}

In this work, we conduct a combined analysis of all available data on cross sections and spin-dependent observables for the $\gamma p \to K^+ \Sigma^0(1385)$, $\gamma n \to K^+ \Sigma^-(1385)$, and $\pi^+ p \to K^+ \Sigma^+(1385)$ reactions within an effective Lagrangian approach. For the background contributions, as depicted in Figs.~\ref{FIG:feymans-g} and \ref{FIG:feymans}, we include the $s$-channel $N$ exchange, $t$-channel $K$ and $K^\ast$ exchanges, $u$-channel $\Lambda$ and $\Sigma^\ast$ exchanges, and the interaction current for the $\gamma p \to K^+ \Sigma^0(1385)$ and $\gamma n \to K^+ \Sigma^-(1385)$ reactions. For the $\pi^+ p \to K^+ \Sigma^+(1385)$ reaction, we consider the $t$-channel $K^\ast$ exchange and $u$-channel $\Lambda$ and $\Sigma$ exchanges. Based on this, we incorporate contributions from the exchanges of $N$ and $\Delta$ resonances in $s$ channel to reproduce the data. Our strategy for selecting resonances is to introduce as few $N$ and $\Delta$ resonances as possible while still achieving a satisfactory fit to the data.

As demonstrated in Ref.~\cite{Wang:2020}, even for the $\gamma p \to K^+ \Sigma^0(1385)$ reaction alone, the differential cross-section data in the center-of-mass energy region $W\le 2200$ MeV cannot be adequately described without including resonance exchanges. In fact, the background contributions are primarily constrained by high-energy data. Specifically, the $t$-channel meson exchanges are well-constrained by high-energy data at forward angles, while the $u$-channel baryon exchanges are largely determined by high-energy data at backward angles. Without considering resonance contributions, the differential cross sections for $\gamma p \to K^+ \Sigma^0(1385)$ are significantly underestimated in the energy region $W\le 2200$ MeV \cite{Wang:2020}, highlighting the necessity of including $N$ or $\Delta$ resonances in this energy range.

\begin{figure}[tbp]
\includegraphics[width=\columnwidth]{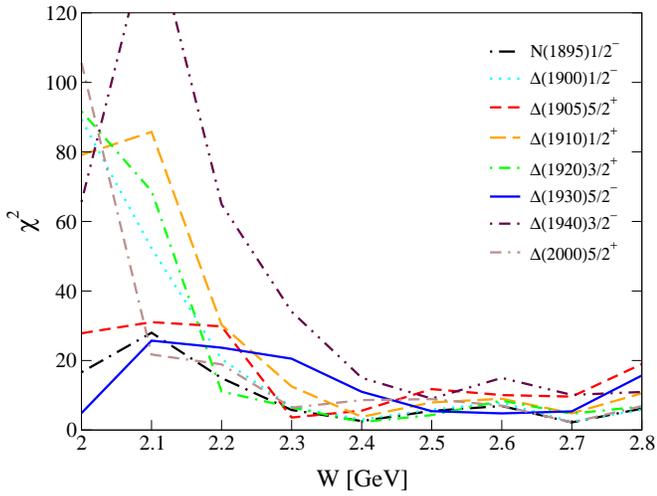}
\caption{The $\chi^2$ values at each center-of-mass energy point for the differential cross sections of $\gamma p \to K^+ \Sigma^0(1385)$ fitted by incorporating contributions from various resonances. }
\label{fig:chi_rp}
\end{figure}

\begin{figure}[tbp]
\includegraphics[width=\columnwidth]{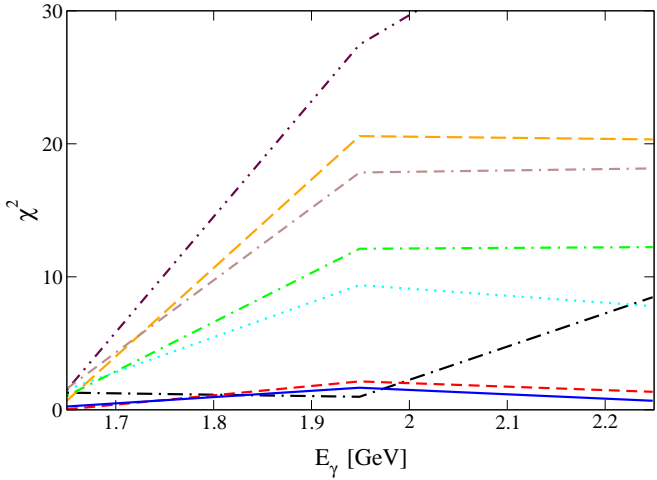}
\caption{The $\chi^2$ values at each data point for the photo-beam asymmetries of $\gamma n \to K^+ \Sigma^-(1385)$ fitted by including contributions from various resonances. Labels for lines are the same as in Fig.~\ref{fig:chi_rp}. }
\label{fig:chi_beam}
\end{figure}

\begin{figure}[tbp]
\includegraphics[width=\columnwidth]{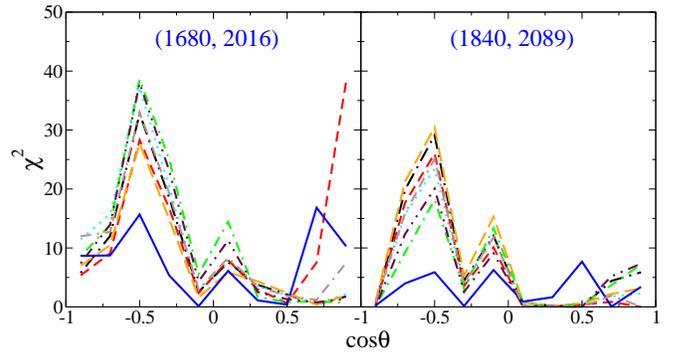}
\caption{The $\chi^2$ values at each data point for the differential cross sections of $\pi^+ p \to K^+ \Sigma^+(1385)$ at selected energies (left and right numbers in parenthesis correspond to photon laboratory energy and center-of-mass energy, respectively) fitted by incorporating contributions from various resonances. Labels for lines are the same as in Fig.~\ref{fig:chi_rp}. }
\label{fig:chi_pi}
\end{figure}

Above the $K\Sigma(1385)$ threshold and below $W=2200$ MeV, there are eight $N$ resonances and six $\Delta$ resonances rated as three-star or four-star states in RPP \cite{PDG:2024}: $N(1875)3/2^-$, $N(1880)1/2^+$, $N(1895)1/2^-$, $N(1900)3/2^+$, $N(2060)5/2^-$, $N(2100)1/2^+$, $N(2120)3/2^-$, $N(2190)7/2^-$, $\Delta(1900)1/2^-$, $\Delta(1905)5/2^+$, $\Delta(1910)1/2^+$, $\Delta(1920)3/2^+$, $\Delta(1930)5/2^-$, and $\Delta(1950)7/2^+$. In addition, there are two $\Delta$ resonances rated as two-star states in RPP \cite{PDG:2024} within this energy range: $\Delta(1940)3/2^-$ and $\Delta(2000)5/2^+$. Note that the $N$ resonances do not contribute to the $\pi^+ p \to K^+ \Sigma^+(1385)$ reaction due to the restriction of isospin conservation. As will be demonstrated and discussed later, the differential cross-section data for $\pi^+ p \to K^+ \Sigma^+(1385)$ clearly indicate the need for resonance contributions. In view of this, we consider the contributions from one of the eight $\Delta$ resonances mentioned above in this study. To illustrate the necessity of $\Delta$ resonance contributions, we also examine a model where resonance contributions are solely from the $N(1895)1/2^-$, which was shown in Ref.~\cite{Wang:2020} to be the only $N$ resonance capable of satisfactorily describing the data for the $\gamma p \to K^+ \Sigma^0(1385)$ reaction. The resonance couplings to $N\pi$ are determined using their partial decay widths to $N\pi$ as quoted in the most recent RPP \cite{PDG:2024}, while their couplings to $K\Sigma(1385)$ and $N\gamma$ are treated as free parameters to be determined by fitting the data.

After systematically testing each of the $\Delta$ resonances and the $N(1895)1/2^-$ resonance individually, we find that only the model incorporating the $\Delta(1930)5/2^-$ resonance is capable of simultaneously describing all available data on differential cross sections and spin-dependent observables for the $\pi^+ p \to K^+ \Sigma^+(1385)$, $\gamma n \to K^+ \Sigma^-(1385)$, and $\gamma p \to K^+ \Sigma^0(1385)$ reactions. In the following discussion, we first explain why models including any resonance other than the $\Delta(1930)5/2^-$ fail to provide acceptable solutions. Subsequently, we present and analyze the results obtained with the inclusion of the $\Delta(1930)5/2^-$ resonance.

\begin{table*}[tbp]
\caption{\label{Table:para} Fitted values of adjustable model parameters for the $\gamma p \to K^+ \Sigma^0(1385)$, $\gamma n \to K^+ \Sigma^-(1385)$, and $\pi^+ p \to K^+ \Sigma^+(1385)$ reactions. The cutoff parameters $\Lambda_t$, $\Lambda_u$, and $\Lambda_s$ are in MeV. The symbol $\Delta^\ast$ denotes $\Delta(1930)5/2^-$. }
\begin{tabular*}{\textwidth}{@{\extracolsep\fill}ccccccccc}
\hline\hline
$A_0$ & $\Lambda_t$ & $\Lambda_u$ & $\Lambda_s$ & $g^{(1)}_{\Sigma^*\Lambda\gamma}/g^{(1)}_{\Sigma^*\Lambda\gamma}$ & $g_{\Delta^\ast N\gamma}^{(1)}$ & $g_{\Delta^\ast N\gamma}^{(2)}$ &    $g_{\Delta^\ast \Sigma^\ast K}^{(1)}$ & $g_{\Delta^\ast \Sigma^\ast K}^{(2)}$  \\
$-0.064\pm 0.004$  &  $860\pm 3$  &  $1132\pm 3$  & $878\pm 4$ &$-0.57\pm0.2$&$0.0\pm 0.2$ &  $-7.90\pm 0.6$  & $-2.22\pm 0.2$ & $-7.71\pm 0.3$ \\
\hline\hline
\end{tabular*}
\end{table*}

\begin{figure}[tbp]
\includegraphics[width=\columnwidth]{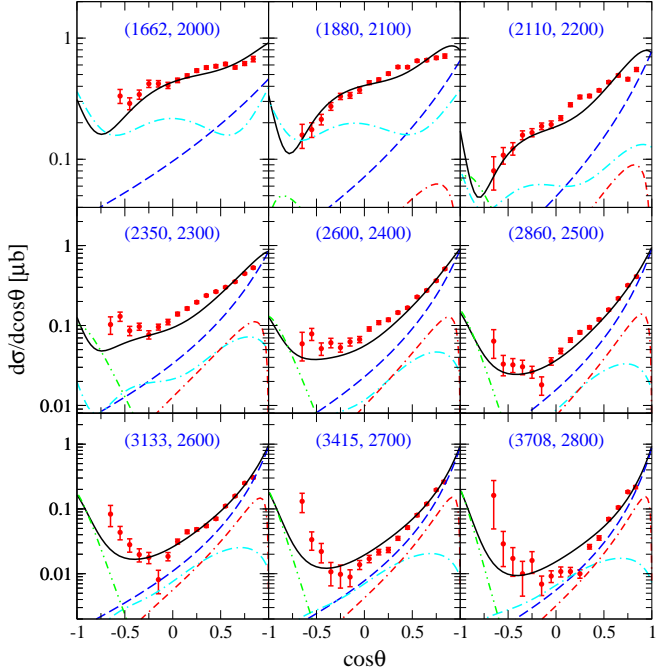}
\caption{Differential cross sections for $\gamma p \to K^+ \Sigma^0(1385)$ as a function of $\cos\theta$ (black solid lines). The red scattered symbols denote the data from the CLAS Collaboration \cite{Mori:2013}. The cyan dash-dotted, blue dashed, red dot-double-dashed, and green dash-double-dotted lines represent the individual contributions from the $s$-channel $\Delta(1930){5/2}^-$ resonance exchange, generalized contact term, $t$-channel $K$ exchange, and $u$-channel $\Lambda$ exchange, respectively. The numbers in parentheses denote the centroid value of the photon laboratory incident energy (left number) and the corresponding total center-of-mass energy of the system (right number), in MeV. }
\label{fig:dif_p}
\end{figure}

\begin{figure}[tbp]
\includegraphics[width=\columnwidth]{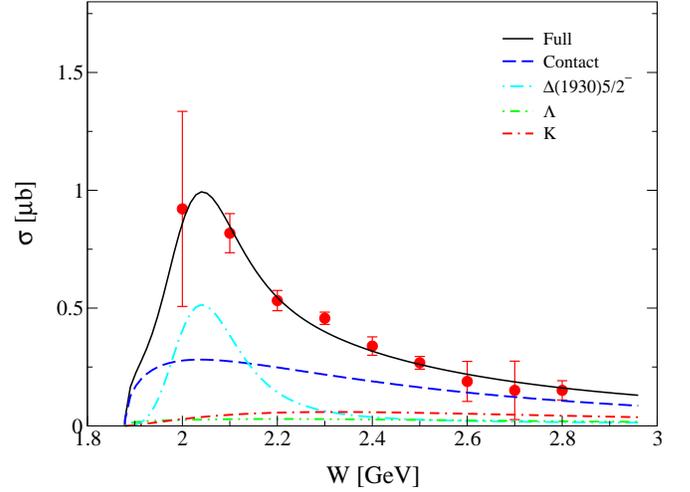}
\caption{Predicted total cross sections for $\gamma p \to K^+ \Sigma^0(1385)$ with dominant individual contributions as a function of center-of-mass incident energy. Data (red full squares) are taken from the CLAS Collaboration \cite{Mori:2013} but not included in the fit.}
\label{fig:total_p}
\end{figure}

\begin{figure}[tbp]
\includegraphics[width=\columnwidth]{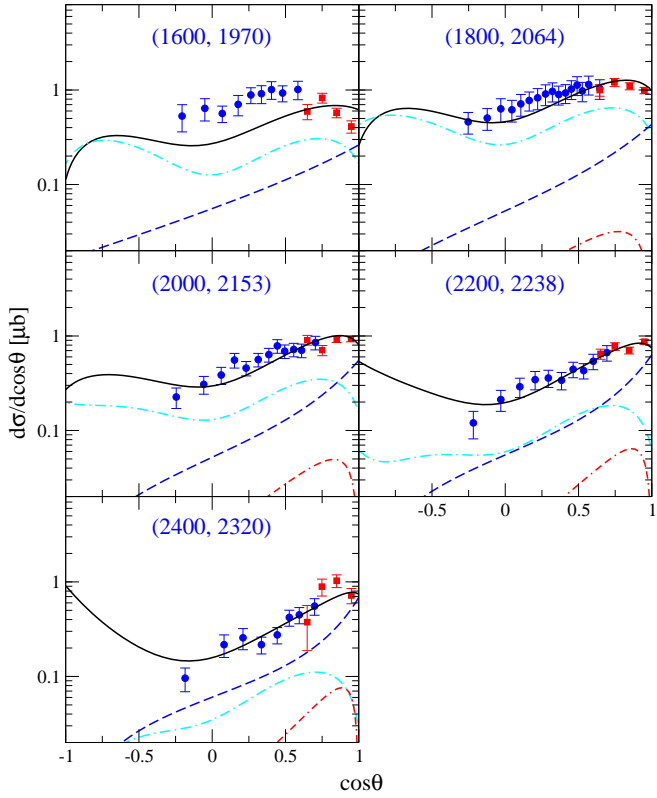}
\caption{Differential cross sections for $\gamma n \to K^+ \Sigma^-(1385)$ as a function of $\cos\theta$ (black solid lines). The red squares and blue circles denote the data from the LEPS Collaboration \cite{Hicks:2009} and CLAS Collaboration \cite{Paul:2014}, respectively. Notations for lines are the same as in Fig.~\ref{fig:dif_p}. }
\label{fig:dif1}
\end{figure}

\begin{figure}[tbp]
\includegraphics[width=\columnwidth]{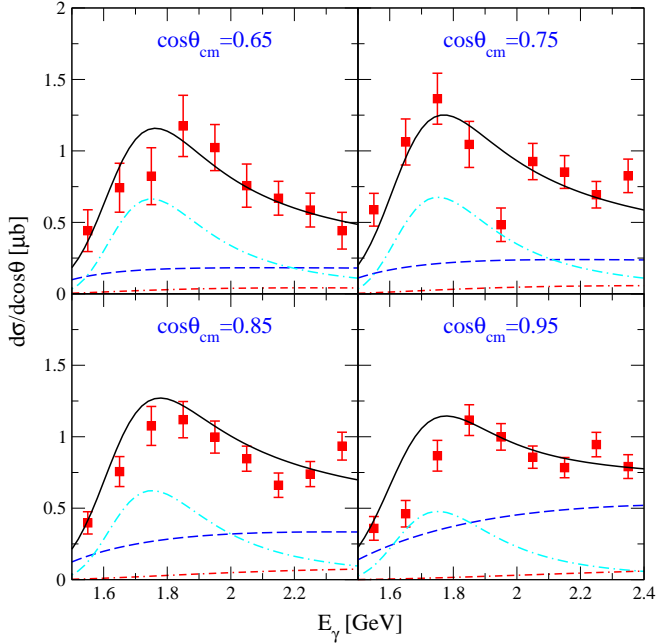}
\caption{Differential cross sections for $\gamma n \to K^+ \Sigma^-(1385)$ as a function of photon incident energy at four $\cos\theta$ intervals (black solid lines). The red full squares denote the data from the LEPS Collaboration \cite{Hicks:2009}. Notations for lines are the same as in Fig.~\ref{fig:dif_p}. }
\label{fig:dif2}
\end{figure}

\begin{figure}[tbp]
\includegraphics[width=\columnwidth]{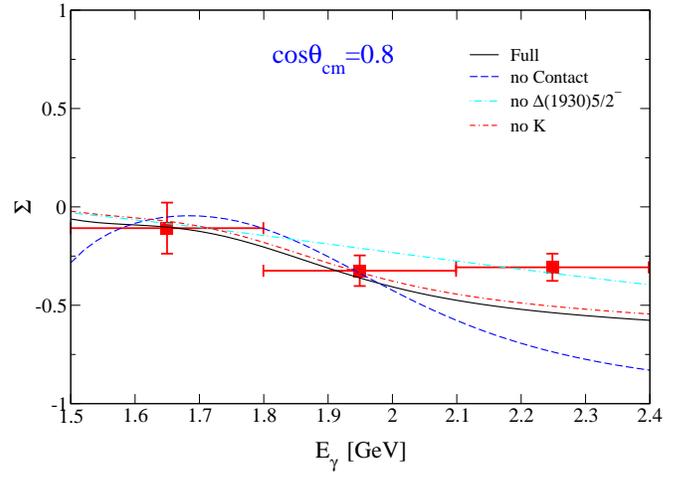}
\caption{Photobeam asymmetries for $\gamma n \to K^+ \Sigma^-(1385)$ plotted against the photon laboratory energy $E_\gamma$. The red full squares denote the data from the LEPS Collaboration \cite{Hicks:2009}. The cyan dash-dotted, blue dashed, and red dot-double-dashed lines represent the results obtained by switching off the individual contributions of the $s$-channel $\Delta(1930){5/2}^-$ exchange, generalized contact term, and $t$-channel $K$ exchange, respectively, from the full reaction amplitudes. Note that the data were measured at $\cos\theta\approx 0.6-1.0$, while the theoretical results are calculated at $\cos\theta=0.8$.}
\label{fig:beam}
\end{figure}

\begin{figure}[tbp]
\includegraphics[width=\columnwidth]{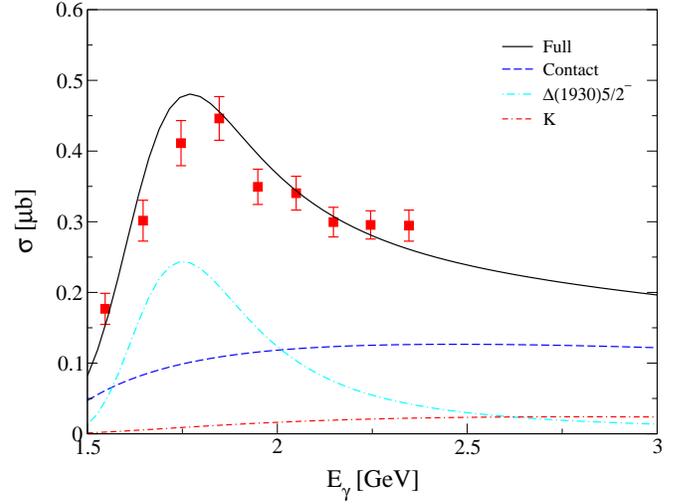}
\caption{Predicted total cross sections for $\gamma n \to K^+ \Sigma^-(1385)$ with dominant individual contributions as a function of photon incident energy. Data (red full squares) are taken from the LEPS Collaboration \cite{Hicks:2009} but not included in the fit.}
\label{fig:total}
\end{figure}

\begin{figure}[tbp]
\includegraphics[width=\columnwidth]{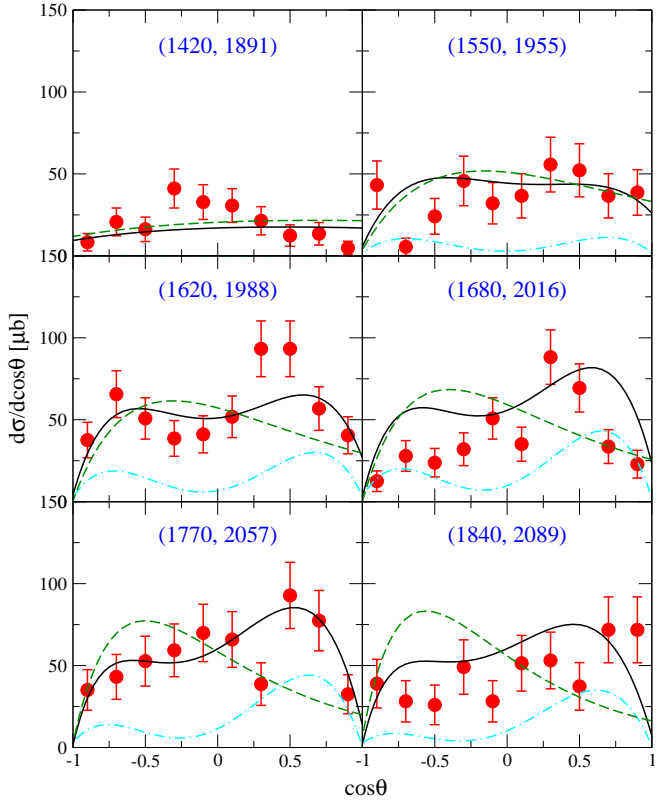}
\caption{Differential cross sections for $\pi^+ p \to K^+ \Sigma^+(1385)$ as a function of $\cos\theta$ (black solid lines). The red circles denote the data taken from Ref.~\cite{Hanson:1972}. The cyan dash-dotted and green dashed lines represent the individual contributions from the $s$-channel $\Delta(1930){5/2}^-$ resonance exchange and $u$-channel $\Lambda$ exchange, respectively. The numbers in parentheses denote the centroid value of the $\pi$ momentum (left number) and the corresponding total center-of-mass energy of the system (right number), in MeV.}
\label{fig:dif_pi}
\end{figure}

\begin{figure}[tbp]
\includegraphics[width=\columnwidth]{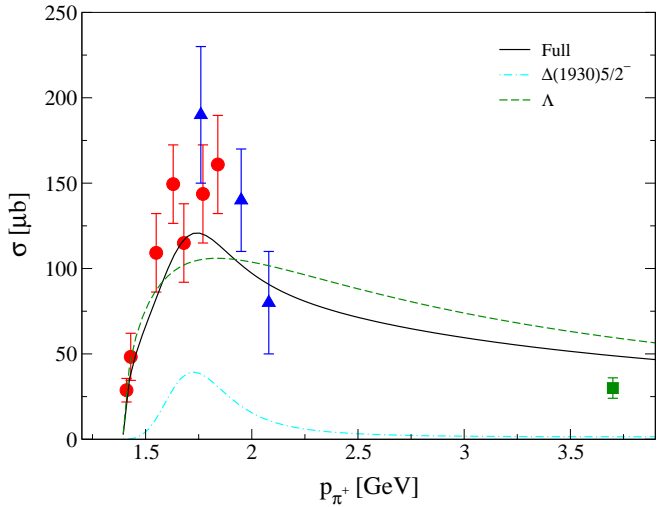}
\caption{Predicted total cross sections for $\pi^+ p \to K^+ \Sigma^+(1385)$ with dominant individual contributions as a function of $\pi$ momentum. Data from Ref.~\cite{Hanson:1972} (red circles), Ref.~\cite{dagan} (blue triangles), and Ref.~\cite{Butler:1973gq} (green squares) are not included in the fit. }
\label{fig:total_pi}
\end{figure}

Figure~\ref{fig:chi_rp} shows the $\chi^2$ values at each center-of-mass energy point for the differential cross sections of $\gamma p \to K^+ \Sigma^0(1385)$ fitted by including one of the $\Delta$ resonances mentioned above or the $N(1895)1/2^-$ resonance\footnote{The $\chi^2$ for the results fitted by including the $\Delta(1950)7/2^+$ is not shown here. As demonstrated in Ref.~\cite{Wang:2020}, the $\chi^2$ for the differential cross-section results of $\gamma p \to K^+ \Sigma^0(1385)$ fitted by including the $\Delta(1950)7/2^+$ resonance is larger than that obtained with the inclusion of the $\Delta(1910)1/2^+$ resonance.}. It is evident that the fits obtained by including any resonance other than the $N(1895)1/2^-$, $\Delta(1905)5/2^+$, or $\Delta(1930)5/2^-$ can be excluded as acceptable, as the corresponding $\chi^2$ values are significantly large at the lowest three energy points. In Fig.~\ref{fig:chi_beam}, we further present the $\chi^2$ values for the photo-beam asymmetries of $\gamma n \to K^+ \Sigma^-(1385)$ at three experimental energy points, fitted by including one of the resonances mentioned above. Obviously, the fit incorporating the $N(1895)1/2^-$ resonance can be ruled out due to the large $\chi^2$ value at the third energy point. Figure~\ref{fig:chi_pi} displays the $\chi^2$ values at each data point for the differential cross sections of $\pi^+p\to K^+ \Sigma^+(1385)$ at the energy points $W=2016$ MeV and $W=2089$ MeV, fitted by including one of the resonances mentioned above. It is apparent that the fit with the $\Delta(1905)5/2^+$ resonance fails to reproduce the differential cross-section data around $\cos\theta \approx -0.5$ at both energy points, as well as the data around $\cos\theta\approx -1$ at $W=2016$ MeV. In a word, when considering the fitting quality for all three reactions, $\gamma p \to K^+ \Sigma^0(1385)$, $\gamma n \to K^+ \Sigma^-(1385)$, and $\pi^+ p \to K^+ \Sigma^+(1385)$, the fit incorporating the $\Delta(1930)5/2^-$ resonance is the only one that provides an acceptable description of the data for all these three reactions. The mass and width of the $\Delta(1930)5/2^-$ resonance, $M_{\Delta(1930)}=1950$ MeV and $\Gamma_{\Delta(1930)}=300$ MeV, are taken from the RPP \cite{PDG:2024}. In the following parts of the paper, we present and discuss the results obtained from this fit.  

Table~\ref{Table:para} presents the fitted values of the adjustable parameters in this study, which enable a satisfactory description of the available data for the three considered reactions. Note that the same cutoff mass $\Lambda_t$ is applied for the $t$-channel $K^\ast$ and $K$ exchanges in the $\gamma p \to K^+ \Sigma^0(1385)$, $\gamma n \to K^+ \Sigma^-(1385)$, and $\pi^+ p \to K^+ \Sigma^+(1385)$ reactions. Similarly, for all these three reactions, the same cutoff mass $\Lambda_u$ is used for the $u$-channel $\Lambda$, $\Sigma$, and $\Sigma^\ast$ exchanges, and the same cutoff mass $\Lambda_s$ is employed for the $s$-channel $N$ and $\Delta(1930)5/2^-$ resonance exchanges. In contrast to Refs.~\cite{Wang:2020,Wang:2022}, where only the product of the electromagnetic and hadronic couplings of the resonance was relevant to the reaction amplitudes and thus make the determination of resonance's individual couplings impossible, in the present work, all related couplings of the resonance can be separately determined. Specifically, as mentioned in Sec.~\ref{Sec:formalism}, the coupling constant of $\Delta(1930)\pi N$ can be firstly fixed by the partial decay width of $\Gamma_{\Delta(1930)\to\pi N} \approx 300 \times 10\% = 30$ MeV \cite{PDG:2024}. As a consequence, the two hadronic couplings $g_{\Delta^\ast \Sigma^\ast K}^{(1)}$ and $g_{\Delta^\ast\Sigma^\ast K}^{(2)}$, as shown in Table~\ref{Table:para}, are determined through fitting the $\pi^+ p \to K^+ \Sigma^+(1385)$ data. Finally, the electromagnetic coupling constants $g_{\Delta^\ast N\gamma}^{(1)}$ and $g_{\Delta^\ast N\gamma}^{(2)}$ are also fixed through fitting the $\gamma p \to K^+ \Sigma^0(1385)$ and $\gamma n \to K^+ \Sigma^-(1385)$ data. It should be noted that the fitting of all these four parameters is not sequential, and the order presented here is just for the purpose of elucidating their logical relationships. By use of the obtained resonance couplings, one can get the helicity amplitudes and branching ratios for the $\Delta(1930)5/2^-$ resonance. The results are as follows: $A^{1/2}=0.054$ GeV$^{-1/2}$, $A^{3/2}=-0.076$ GeV$^{-1/2}$, ${\rm Br}[\Delta(1930)\to N\gamma]=0.084\%$, and ${\rm Br}[\Delta(1930)\to K\Sigma(1385)]=0.53\%$. In the RPP \cite{PDG:2024}, there are no advocated values for these quantities.

The theoretical results for the differential cross sections of the reaction $\gamma p \to K^+ \Sigma^0(1385)$ are presented in Fig.~\ref{fig:dif_p}. In this figure, the cyan dash-dotted, blue dashed, red dot-double-dashed, and green dash-double-dotted lines represent the individual contributions from the $s$-channel $\Delta(1930){5/2}^-$ resonance exchange, generalized contact term, $t$-channel $K$ exchange, and $u$-channel $\Lambda$ exchange, respectively. As shown, the differential cross-section data for this reaction are well reproduced. The $\Delta(1930){5/2}^-$ resonance exchange and the interaction current dominate the differential cross sections in the low-energy region. In the high-energy region, the interaction current, followed by the $t$-channel $K$ exchange, dominates the differential cross sections at forward angles, while the $u$-channel $\Lambda$ exchange dominates the differential cross sections at backward angles. Compared to the results in Ref.~\cite{Wang:2020}, the $\Delta(1930){5/2}^-$ exchange provides larger contributions in the present work, whereas the $K$ exchange and interaction current contribute less. Note that the coupling $g_{\Delta^\ast \Sigma^\ast K}^{(2)}$ was set to be zero in Ref.~\cite{Wang:2020} to reduce the number of parameters, whereas it is treated as a fit parameter in the present work due to the inclusion of a significantly larger number of data points in the fit. The smaller value of the cutoff parameter $\Lambda_t$ in the present work explains why the contributions from the $t$-channel $K$ exchange are smaller than those in Ref.~\cite{Wang:2020}.
 
In Fig.~\ref{fig:total_p}, we present our predictions for the total cross sections of the reaction $\gamma p \to K^+ \Sigma^0(1385)$, along with the dominant individual contributions, as a function of center-of-mass incident energy. One sees that the data are well reproduced. Note that these data were not included in the fitting process. It is evident that the interaction current dominates the background contributions across the entire energy range considered, while the $s$-channel $\Delta(1930){5/2}^-$ resonance exchange accounts for the bump structure observed in the data near $W\approx 2$ GeV. Additionally, the $t$-channel $K$ exchange and the $u$-channel $\Lambda$ exchange also make non-negligible contributions to this reaction.

The theoretical results for the differential cross sections and photobeam asymmetries of the $\gamma n \to K^+ \Sigma^-(1385)$ reaction are shown in Figs.~\ref{fig:dif1}-\ref{fig:beam}. In these figures, the black solid lines represent the outcomes from the full calculation. In Figs.~\ref{fig:dif1} and \ref{fig:dif2}, the cyan dash-dotted, blue dashed, and red dot-double-dashed lines represent the individual contributions from the $s$-channel $\Delta(1930){5/2}^-$ resonance exchange, interaction current, and $t$-channel $K$ exchange, respectively. Contributions from other terms are relatively small and are not clearly visible on the given scale, thus are omitted from the plots. In Fig.~\ref{fig:beam}, the cyan dash-dotted, blue dashed, and red dot-double-dashed lines represent the results obtained by switching off the individual contributions of the $s$-channel $\Delta(1930){5/2}^-$ exchange, interaction current, and $t$-channel $K$ exchange, respectively, from the full photoproduction amplitudes. These figures demonstrate that all available data on differential cross sections and photobeam asymmetries for the $\gamma n \to K^+\Sigma^-(1385)$ reaction are well reproduced. The $\Delta(1930){5/2}^-$ resonance exchange provides the dominant contributions to the cross sections in the low-energy region, while the interaction current also contributes significantly across the entire energy range. In Fig.~\ref{fig:dif1}, it is observed that the CLAS differential cross-section data \cite{Paul:2014} at the lowest energy are somewhat underestimated. Note that the experimental data were measured at $E_\gamma = 1500-1700$ MeV, whereas our theoretical results are calculated at $E_\gamma=1600$ MeV. In the near-threshold energy region, the phase space is highly sensitive to the energy, which may partially explain the discrepancy observed at the lowest energy point in Fig.~\ref{fig:dif1}, as discussed in Ref.~\cite{Wang:2022}.

In Fig.~\ref{fig:total}, the predicted total cross sections for $\gamma n\to K^+\Sigma^-(1385)$, along with the dominant individual contributions, are presented as a function of photon incident energy. Note that the LEPS data were measured in the angular interval $\cos\theta \approx 0.6-1.0$, and correspondingly, our theoretical total cross sections are calculated by integrating the differential cross sections over this angular interval. One sees that the predicted total cross sections qualitatively reproduce the experimental data. The $\Delta(1930){5/2}^-$ resonance exchange dominates the cross sections in low-energy region and is responsible for the bump structure observed near $E_\gamma\approx 1.75$ GeV. The contact term provides significant contributions across the whole energy region considered. Considerable contributions are also seen from the $K$ exchange. 

The theoretical results for the differential cross sections of the $\pi^+ p \to K^+ \Sigma^+(1385)$ reaction are presented in Fig.~\ref{fig:dif_pi}. There, the cyan dash-dotted and green dashed lines represent the individual contributions from the $s$-channel $\Delta(1930){5/2}^-$ resonance exchange and $u$-channel $\Lambda$ exchange, respectively. One sees that our theoretical calculations successfully reproduce the differential cross-section data for the $\pi^+ p \to K^+ \Sigma^+(1385)$ reaction \cite{Hanson:1972}. The dominant contributions are from the $u$-channel $\Lambda$ exchange, while significant contributions are also observed from the $s$-channel $\Delta(1930){5/2}^-$ exchange, which accounts for the double bump structures in the angular distributions.   

In the literature, a comprehensive analysis of the $\pi^+ p \to K^+ \Sigma^+(1385)$ and $pp \to nK^+\Sigma^+(1385)$ reactions was conducted in Ref.~\cite{Xie:2014}. The study suggested that the resonance $\Delta(1940)3/2^-$, with a mass near $1940$ MeV and a width of approximately $200$ MeV, plays a dominant role in the $\pi^+ p \to K^+ \Sigma^+(1385)$ reaction. However, the differential cross-section data for the $\pi^+ p \to K^+ \Sigma^+(1385)$ reaction \cite{Hanson:1972} were not well reproduced across most energy ranges in Ref.~\cite{Xie:2014}. 

The description of the differential cross-section data for $\pi^+ p \to K^+ \Sigma^+(1385)$ in the present work, as shown in Fig.~\ref{fig:dif_pi}, is significantly improved compared to that in Ref.~\cite{Xie:2014}. In particular, the double bump structure in the angular distributions at $P_{\pi}=1620$ MeV is reproduced only in the present work and not in Ref.~\cite{Xie:2014}. This improvement can be attributed to the inclusion of the $\Delta(1930){5/2}^-$ resonance, which itself exhibits two bump structures in its individual angular distributions, as shown by the cyan dash-dotted line in Fig.~\ref{fig:dif_pi}. In contrast, the $\Delta(1940)3/2^-$ resonance included in Ref.~\cite{Xie:2014} results in angular distributions that differ significantly from the data.

In Fig.~\ref{fig:total_pi}, our predictions for the total cross sections of $\pi^+ p \to K^+ \Sigma^+(1385)$, along with the dominant individual contributions, are presented as a function of the $\pi$ momentum. One sees that the data are qualitatively reproduced, even though they were not considered in the fitting process. Notably, the $u$-channel $\Lambda$ exchange dominates the background contributions across the entire energy range considered, while the $s$-channel $\Delta(1930){5/2}^-$ resonance is responsible for the bump structure observed near the $K\Sigma(1385)$ threshold region.

Finally, we note that introducing additional resonances into the model could, in principle, improve the theoretical description of the data, as this would increase the number of adjustable parameters. However, given the limited amount of available data--some of which have large uncertainties--and considering that the current theoretical model already describes the data satisfactorily, we postpone the inclusion of additional resonances until more data, particularly high-precision measurements, become available in the future.

\section{Summary and conclusion}  \label{Sec:summary}

In this study, we perform a comprehensive analysis of all available data on differential cross sections and spin-dependent observables for the $\gamma p \to K^+ \Sigma^0(1385)$, $\gamma n \to K^+ \Sigma^-(1385)$, and $\pi^+ p \to K^+ \Sigma^+(1385)$ reactions using an effective Lagrangian approach. For the $\gamma p \to K^+ \Sigma^0(1385)$ and $\gamma n \to K^+ \Sigma^-(1385)$ reactions, the background contributions are constructed by incorporating $t$-channel exchanges of $K$ and $K^\ast(892)$, $s$-channel contributions from $N$, $u$-channel exchanges of $\Lambda$ and $\Sigma^\ast(1385)$, and an interaction current introduced to ensure the gauge invariance of the full photoproduction amplitudes. For the $\pi^+ p \to K^+ \Sigma^+(1385)$ reaction, the background contributions are modeled using $t$-channel $K^\ast(892)$ exchange and $u$-channel $\Lambda$ and $\Sigma$ exchanges. Regarding  resonance contributions, we introduce the minimal number of $N$ and $\Delta$ resonances required to reproduce the experimental data. After extensive testing, we find that all available data for those three reactions can be satisfactorily described within a unified theoretical framework by including the $\Delta(1930){5/2}^-$ resonance in the $s$ channel. Other near-threshold resonances, such as $N(1895)1/2^-$, $\Delta(1900)1/2^-$, $\Delta(1905)5/2^+$, $\Delta(1910)1/2^+$, $\Delta(1920)3/2^+$, $\Delta(1940)3/2^-$, and $\Delta(2000)5/2^+$, were individually tested but failed to reproduce the data. 

For the $\gamma p \to K^+ \Sigma^0(1385)$ reaction, the $\Delta(1930){5/2}^-$ resonance plays a crucial role in reproducing the angular distribution structures exhibited in the differential cross-section data near the threshold energy region. The contact term dominates the background contributions, while the $K$ and $\Lambda$ exchanges make considerable contributions across various angular regions. In the case of the $\gamma n \to K^+ \Sigma^-(1385)$ reaction, the interaction current dominates the background contributions across the entire energy range considered, with the $\Delta(1930){5/2}^-$ resonance and $K$ exchanges also providing substantial contributions. For the $\pi^+ p \to K^+ \Sigma^+(1385)$ reaction, both the $s$-channel $\Delta(1930){5/2}^-$ and $u$-channel $\Lambda$ exchanges are dominant. Notably, the $\Delta(1930){5/2}^-$ exchange accounts for the double-peak structure observed in the angular distribution data around $P_{\pi}=1620$ MeV. Compared to the results of Ref.~\cite{Xie:2014}, where the $\Delta(1940)3/2^-$ was proposed to dominate the cross sections of $\pi^+ p \to K^+ \Sigma^+(1385)$, the theoretical description of the $\pi^+ p \to K^+ \Sigma^+(1385)$ data in this work is significantly improved. This suggests that the $\Delta(1930){5/2}^-$ is more appropriate than the $\Delta(1940)3/2^-$ for describing the data of the $\pi^+ p \to K^+ \Sigma^+(1385)$ reaction.

To further elucidate the reaction mechanisms of the $\gamma p \to K^+ \Sigma^0(1385)$, $\gamma n \to K^+ \Sigma^-(1385)$, and $\pi^+ p \to K^+ \Sigma^+(1385)$ reactions, and to more reliably extract the resonance contents and parameters in these reactions, future experiments are encouraged to provide additional data, particularly on spin observables, for these reactions.

\begin{acknowledgments}
This work is partially supported by the National Natural Science Foundation of China (Grants No.~12305097, No.~12175240, No.~12147153, and No.~12305137), the Fundamental Research Funds for the Central Universities, the Taishan Scholar Young Talent Program (Grant No.~tsqn202408091), and the Shandong Provincial Natural Science Foundation, China (Grant No.~ZR2024QA096).
\end{acknowledgments}


\begin{thebibliography}{99}
%
\bibitem{Isgur:1978}
N. Isgur and G. Karl, Phys. Rev. D {\bf 18}, 4187 (1978).
%
\bibitem{Capstick:1986}
S. Capstick and N. Isgur, Phys. Rev. D {\bf 34}, 2809 (1986).
%
\bibitem{Loring:2001}
U. L\"{o}ring, B. C. Metsch, and H. R. Petry, Eur. Phys. J. A {\bf 10}, 395 (2001).
%
\bibitem{Mori:2013}
K. Moriya {\it et al.}  (CLAS Collaboration), Phys. Rev. C {\bf 88}, 045201 (2013).
%
\bibitem{Hicks:2009}
K. Hicks {\it et al.} (LEPS Collaboration), Phys. Rev. Lett. {\bf 102}, 012501 (2009).
%
\bibitem{Paul:2014}
Paul Mattione (CLAS Collaboration), Int. J. Mod. Phys. Conf. Ser. {\bf 26}, 1460101 (2014).
%
\bibitem{Hanson:1972}
P. Hanson, G. E. Kalmus, and J. Louie, Phys. Rev. D {\bf 4}, 1296 (1971).
%
\bibitem{Wang:2020}
A. C. Wang, W. L. Wang, and F. Huang, Phys. Rev. D {\bf 101}, 074025 (2020).
%
\bibitem{Wang:2022}
A. C. Wang, N. C. Wei, and F. Huang, Phys. Rev. D {\bf 105}, 034017 (2022).
%
\bibitem{Haberzettl:1997}
H. Haberzettl, Phys. Rev. C {\bf 56}, 2041 (1997).
%
\bibitem{Haberzettl:2006}
H. Haberzettl, K. Nakayama, and S. Krewald, Phys. Rev. C {\bf 74}, 045202 (2006).
%
\bibitem{Huang:2012}
F. Huang, M. D\"{o}ring, H. Haberzettl, J. Haidenbauer, C. Hanhart, S. Krewald, U.-G. Mei{\ss}ner, and K. Nakayama, Phys. Rev. C {\bf 85}, 054003 (2012).
%
\bibitem{Huang:2013}
F. Huang, H. Haberzettl, and K. Nakayama, Phys. Rev. C {\bf 87}, 054004 (2013).
%
\bibitem{Drell:1972}
S. D. Drell and T. D. Lee, Phys. Rev. D {\bf 5}, 1738 (1972).
%
\bibitem{PDG:2024}
S. Navas {\it et al.} (Particle Data Group), Phys. Rev. D {\bf 110}, 030001 (2024).
%
\bibitem{Lichtenberg:1977}
D. B. Lichtenberg, Phys. Rev. D {\bf 15}, 345 (1977).
%
\bibitem{Garcilazo:1993}
H. Garcilazo and E. Moya de Guerra, Nucl. Phys. {\bf A562}, 521 (1993).
%
\bibitem{Behrends:1957}
R. E. Behrends and C. Fronsdal, Phys. Rev. {\bf 106}, 345 (1957).
%
\bibitem{Fronsdal:1958}
C. Fronsdal, Suppl. Nuovo Cimento {\bf 9}, 416 (1958).
%
\bibitem{Zhu:1999}
J. J. Zhu and M. L. Yan, arXiv:hep-ph/9903349.
%
\bibitem{Kim:2014}
S. H. Kim, A. Hosaka, and H. C. Kim, Phys. Rev. D {\bf 90}, 014021 (2014).
%
\bibitem{Wang:2017}
A. C. Wang, W. L. Wang, F. Huang, H. Haberzettl, and K. Nakayama, Phys. Rev. C {\bf 96}, 035206 (2017).
%
\bibitem{Wang:2018}
A. C. Wang, W. L. Wang, and F. Huang, Phys. Rev. C {\bf 98}, 045209 (2018).
%
\bibitem{Kim:2011}
S. H. Kim, S. I. Nam, Y. Oh, and H. C. Kim, Phys. Rev. D {\bf 84}, 114023 (2011).
%
\bibitem{Xie:2014}
J. J. Xie, E. Wang, and B. S. Zou, Phys. Rev. C {\bf 90}, 025207 (2014).
%
\bibitem{dagan}
S. Dagan, Z. M. Ma, J. W. Chapman, L. R. Fortney, and E. C. Fowler, Phys. Rev. {\bf 161}, 1384 (1967).
%
\bibitem{Butler:1973gq}
W. R. Butler, D. G. Coyne, G. Goldhaber, J. Macnaughton, and G. H. Trilling, Phys. Rev. D {\bf 7}, 3177 (1973).
%
\end{thebibliography}
\end{document}